%% file: parallel_tacotron.tex
\documentclass{article}
\usepackage[preprint]{spconf}
\usepackage{amsmath,graphicx}
\usepackage{hyperref}
\usepackage{booktabs}
\usepackage{wrapfig}
\usepackage{amssymb}
\usepackage{bm}
\usepackage{url}
\usepackage{float}
\usepackage{tikz}
\usepackage{pifont}
\usepackage{microtype}
\usepackage{cite}
\usepackage{newtxtext,newtxmath}

\title{Parallel Tacotron: Non-Autoregressive and Controllable TTS}

\name{Isaac~Elias,~~Heiga~Zen,~~Jonathan~Shen,~~Yu~Zhang,~~Ye~Jia,~~Ron~J.~Weiss,~~Yonghui~Wu}
\address{
    Google
}

\begin{document}
\ninept
\maketitle

\begin{abstract}
\input{abstract}
\end{abstract}
 \copyrightnotice{\begin{tabular}[t]{@{}l@{}}\copyright IEEE 2020 Personal use of this material is permitted. Permission from IEEE must be obtained for all other uses, in any \\current or future media, including reprinting/republishing this  material for advertising or promotional purposes, creating new \\collective works, for resale or redistribution to servers or lists, or reuse of any copyrighted component of this work in other works.\end{tabular}}

\section{Introduction}
\label{sec:intro}
\input{introduction.tex}

\section{Parallel Tacotron}
\label{sec:architecture}

\begin{figure*}[t]
\centering
\includegraphics[width=0.8\textwidth]{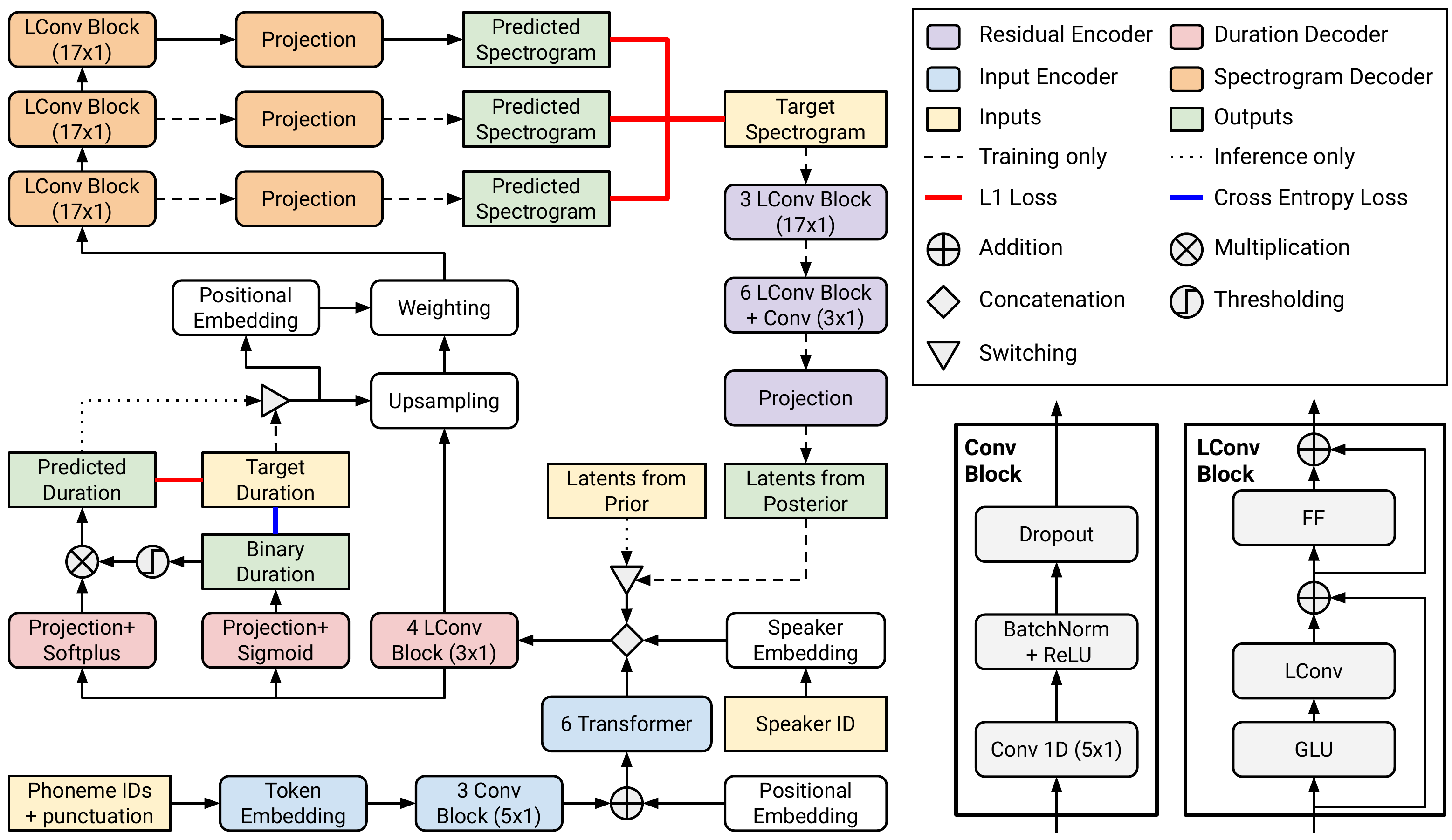}
\caption{Block diagram of the Parallel Tacotron model.  The residual encoder (purple blocks) in this figure correspond to the global VAE in Section~\ref{sec:global_vae}.}
\label{fig:architecture}
\end{figure*}

Figure~\ref{fig:architecture} illustrates the architecture of the Parallel Tacotron model.
It consists of an  input encoder,  a residual encoder,  a duration decoder, an upsampling block, and  a spectrogram decoder stack.

The model heavily relies on self-attention blocks with either Transformer or lightweight convolutions (LConv).
The LConv is a depth-wise convolution which shares certain output channels whose weights are normalized across time \cite{wu2019pay}.
It has a few orders of magnitude less parameters than a standard non-separable convolution.
Unlike Transformer-based self-attention, lightweight convolutions have a fixed context window and reuse the same weights for context elements, regardless of the current time-step.
This can be more useful in TTS as relevant context elements can be more local than machine translation or other language tasks.
The LConv block is also illustrated in Fig.~\ref{fig:architecture}.
It consists of a gated linear unit (GLU), a lightweight convolution, and a feedforward (FF) layer with residual connections.
As in \cite{wu2019pay} we use dropout 0.1, and perform FF mixing in the FF layer using the structure $\text{ReLU}(\bm{W}_1\bm{X} +\bm{b_1})\bm{W}_2 +\bm{b}_2$ where $\bm{W}_1$ increases the dimension by a factor 4.

\subsection{Input Encoder}
As in Tacotron 2 \cite{tacotron2}, the input encoder consists of a phoneme embedding lookup followed by three convolutional blocks.
Thereafter, sinusoidal positional embeddings are added, then six Transformer blocks are applied.\footnote{An LConv stack works equally well.}  The output of the encoder can be viewed as a contextual phoneme representation.
The encoder outputs are concatenated with a speaker embedding of dimension 64 and latent representation from the residual encoder.

\subsection{Variational Residual Encoder}
Two different VAE models are investigated; a global VAE per speaker similar to \cite{hsu2018hierarchical} and a fine-grained phoneme-level VAE akin to \cite{sun2020fullyhierarchical}.

\subsubsection{Global VAE per Speaker}
\label{sec:global_vae}

Like multi-speaker modeling in \cite{hsu2018hierarchical}, the model is augmented by a global VAE per speaker, each with a speaker specific learned prior.
The latent representation from the VAE can be viewed as the residual information which cannot be represented by input phoneme sequences and associated speaker IDs, such as prosodic variations \cite{hsu2018hierarchical}.
The residual encoder, also known as the posterior network, takes target mel-spectrograms as its input and processes it using a stack of LConv blocks. The stack consists of; three $17 \times 1$ LConv blocks, followed by five $17 \times 1$ LConv blocks interleaved with strided $3 \times 1$ convolutions, where all LConv blocks have eight heads.
This lets the model successively downsample the representation before applying global average pooling to get the final global latent representation.
Although the dimension of the latent representation is 8, it is projected up to 32. 
During inference, a speaker specific prior mean is used.

In addition to modeling the residual information, this latent representation works to implicitly capture the dependencies among output mel-spectrogram frames given input tokens \cite{gu2017,lee-arxiv-2018,shu-aaai-2020}.

\subsubsection{Phoneme-Level Fine-Grained VAE}
A fine-grained phoneme-level VAE \cite{sun2020fullyhierarchical} has the posterior network which aligns the groundtruth spectrogram with the encoder outputs using attention. 
To achieve this, the positional embeddings from Section~\ref{sec:positional} and the speaker embedding are concatenated to the groundtruth spectrogram frames, subsequently five 8-headed $17 \times 1$ LConv blocks are applied.  The attention is then computed with the layer normalized \cite{ba2016layer} encoder outputs. We use an 8-dimensional latent representation, concatenate the speaker embedding and encoder outputs, and then project it to dimension 32.

Furthermore, similar to \cite{sun2020fullyhierarchical}, we train a separate autoregressive LSTM to predict a sequence of latent representation during inference. 
Similarly with the posterior network, this LSTM takes both speaker embedding and encoder outputs as its input.  
The LSTM is trained to predict the posterior mean using the $L_2$ loss with teacher forcing. This loss is computed independently of the posterior network; no error gradient is back-propagated to the posterior network.

\subsection{Duration Decoder}
The encoder outputs augmented by the latent representation and speaker embeddings are fed to a duration decoder which predicts phoneme duration.
We assume that groundtruth phoneme durations are provided by an external aligner, such as hidden Markov model (HMM)-based one \cite{talkin1994aligner}.
Here the input sequence consists of regular phonemes, silences at all word boundaries, and punctuation marks.
Whereas regular phonemes usually have non-zero duration, zero-length duration are often associated with silences and punctuation marks (\textit{e.g.}, word boundaries without pauses).

To model such sequences, the duration decoder predicts two types of outputs: binary zero/non-zero duration and continuous phoneme duration in seconds.
The duration decoder contains four LConv blocks with 3 $\times$ 1 lightweight convolutions followed by two independent projections.
One is followed by a sigmoid activation predicting the probability of non-zero duration $p_z$.
Another is followed by a softplus activation predicting the phoneme duration in seconds.
If $p_z<0.99$ then the predicted duration is zeroed out. 
Here the cross-entropy (CE) and $L_1$ loss functions are used for the former and latter, respectively.

\subsection{Upsampling and Positional Embeddings}
\label{sec:positional}
Upsampling takes the activation of the last LConv block in the duration decoder and upsamples them to the length of target spectrogram frame sequence using the phoneme duration.  
After upsampling, three different types of positional embeddings are added to make the spectrogram decoder aware of the frame positions within phonemes:
(1) Transformer-style sinusoidal embedding \cite{transformer} of a frame position within a phoneme, (2) Transformer-style sinusoidal embedding \cite{transformer} of phoneme duration, (3) Fractional progression of a frame in a phoneme (1D CoordConv \cite{liu-neurips-2018}).

Since several positional embeddings are added, per channel a learned weighted sum is used to combine them with the output of the duration decoder.
Specifically, the weights are softmax normalized per channel; which lets the network select for each channel which embedding that is preferred by the spectrogram decoder.

\subsection{Spectrogram Decoder with Iterative Loss}
\label{iterative_loss}

The output of the upsampling stack is fed to the spectrogram decoder.
It has six 8-headed self-attention blocks with 17 $\times$ 1 LConv with dropout 0.1 as in \cite{wu2019pay}. 
The output of the last FF layer in each block is then projected to 128-bin mel-spectrogram.

We investigate the use of an iterative loss and its impact on the naturalness of synthesized speech;
the outputs of the last FF layers in the self-attention blocks are independently projected to the mel-spectrogram then $L_1$ losses between the predicted and target mel-spectrogram are summed to form the final loss. 
This type of iterative loss has previously been used in \cite{lee-arxiv-2018,dejavu}.

\subsection{Training Objective}
The overall loss functions for the Parallel Tacotron models with global and fine-grained VAEs become
\begin{align}
    \mathcal{L}_\text{global} &= \frac{1}{KT} \sum_i{\mathcal{L}_{\text{spec}_i}} + \frac{1}{N} \lambda_{\text{dur}} \mathcal{L}_{\text{dur}} - \beta D_\mathrm{KL}, \\
     \mathcal{L}_\text{phone} &= \frac{1}{KT} \sum_i{\mathcal{L}_{\text{spec}_i}} + \frac{1}{N} \lambda_{\text{dur}} \mathcal{L}_{\text{dur}} - \beta D_\mathrm{KL} + \frac{1}{N} \mathcal{L}_{\text{prior}},
\end{align}
where $\mathcal{L}_{\text{spec}_i}$ is the $L_1$ spectrogram loss for the $i$-th LConv block in the spectrogram decoder, $\mathcal{L}_{\text{dur}}$ is the two-level duration loss, $D_\mathrm{KL}$ is the KL divergence between prior and posterior from the residual encoder, $\mathcal{L}_{\text{prior}}$ is the learned prior loss in the phoneme-level VAE, $T$ is the total number of frames, $K$ is the size of spectrogram, and $N$ is the number of tokens.

\section{Experiments}
\label{sec:experiments}
\input{experiments.tex}

\section{Conclusions}
\label{sec:conclusion}
A non-autoregressive neural TTS model called Parallel Tacotron was proposed.
It matched the baseline Tacotron~2 in naturalness and offered significantly faster inference than Tacotron~2.
We also showed that both variational residual encoders and an iterative loss improved the naturalness, and the use of lightweight convolutions as self-attention improved both naturalness and efficiency.

Future work includes the direct comparison against duration-based models with the autoregressive decoder and those using additional representation such as $F_0$ and energy.
Investigating better fine-grained variational models is also necessary. 

\section{Acknowledgements}

The authors would like to thank Mike Chrzanowski, Tejas Iyer, Vincent Wan, and Norman Casagrande for their help.
\clearpage

\footnotesize
\bibliographystyle{IEEEbib}
\bibliography{parallel_tacotron}

\end{document}

%% file: abstract.tex
Although neural end-to-end text-to-speech models can synthesize highly natural speech, there is still room for improvements to its efficiency and naturalness.
This paper proposes a non-autoregressive neural text-to-speech model augmented with a variational autoencoder-based residual encoder.
This model, called \emph{Parallel Tacotron}, is highly parallelizable during both training and inference, allowing efficient synthesis on modern parallel hardware.
The use of the variational autoencoder relaxes the one-to-many mapping nature of the text-to-speech problem and improves naturalness.
To further improve the naturalness, we use lightweight convolutions, which can efficiently capture local contexts, and introduce an iterative spectrogram loss inspired by iterative refinement. 
Experimental results show that Parallel Tacotron matches a strong autoregressive baseline in subjective evaluations with significantly decreased inference time.

%% file: introduction.tex
Neural end-to-end text-to-speech (TTS) has been researched extensively in the last few years \cite{char2wav,tacotron,deepvoice,li-aaai-2019}. 
Tacotron 2, which combined an attention-based encoder-decoder model predicting a mel-spectrogram given a character sequence and a WaveNet model \cite{oord-arxiv-2016} predicting speech waveform given the predicted mel-spectrogram, could synthesize read speech almost as natural as human \cite{tacotron2}.
Following the same approach, a number of models have been proposed to synthesize various aspects of speech (\textit{e.g.}, speaking styles) in a natural sounding way.

Tacotron 2 uses an autoregressive uni-directional long short-term memory (LSTM)-based decoder with the soft attention mechanism \cite{bahdanau2014neural}.
This architecture makes both training and inference less efficient on modern parallel hardware 
than fully feed-forward architectures.
Furthermore, like other autoregressive models, Tacotron 2 uses teacher forcing \cite{teacherforcing}, which introduces discrepancy between training and inference \cite{scheduled_sampling,professor_forcing}. 
Together with the soft attention mechanism, it can lead to robustness errors such as babbling, early cut-off, word repetition, and word skipping
\cite{he2019robust,zheng-taslp-2019,guo-interspeech-2019,battenberg-icassp-2020}.
Although Transformer TTS \cite{li-aaai-2019} addresses the inefficiency during training, it is still inefficient during inference and has the potential for robustness errors due to the autoregressive decoder and the attention mechanism.

Recently a number of neural TTS models relying on explicit phoneme duration rather than the soft attention mechanism were proposed \cite{ren2019fastspeech,ren2020fastspeech2,beliaev-arxiv-2020,lim-arxiv-2020,zeng-icassp-2020,miao-icassp-2020,donahue-arxiv-2020,lancucki-arxiv-2020,yu2019durian,shen-arxiv-2020}.
They are analogous to conventional parametric synthesis models \cite{Zen_SPSS_SPECOM,zen-icassp-2013}, where phoneme durations are often modeled explicitly.
As these models are based on phoneme durations, they are less prone to synthesize speech with the robustness errors.
Furthermore, rhythm of synthesized speech can be controlled precisely by modifying predicted durations.
Some of these models train their explicit duration model to predict groundtruth phoneme duration extracted from autoregressive models \cite{ren2019fastspeech,lim-arxiv-2020,lancucki-arxiv-2020}, internal aligner models \cite{beliaev-arxiv-2020,zeng-icassp-2020}, or external aligner models \cite{ren2020fastspeech2,yu2019durian,shen-arxiv-2020}. 
Among these duration-based models, some of them use the autoregressive decoder \cite{yu2019durian,shen-arxiv-2020}, whereas others use a non-autoregressive decoder \cite{ren2019fastspeech,ren2020fastspeech2,beliaev-arxiv-2020,lim-arxiv-2020,zeng-icassp-2020,miao-icassp-2020,donahue-arxiv-2020,lancucki-arxiv-2020}, which are more efficient during both training and inference.

Text-to-speech synthesis is a one-to-many mapping problem, as there can be multiple possible speech realizations with different prosody for a text input. 
Neural TTS models with autoregressive decoders can use the previous mel-spectrogram frame and the text to predict the next mel-spectrogram frame, where the previous frame can provide context to disambiguate between multi-modal outputs.
However, models with non-autoregressive decoders need a different way to get context to choose a mode.
FastSpeech \cite{ren2019fastspeech} used knowledge distillation to handle this issue, whereas
FastSpeech 2 \cite{ren2020fastspeech2} addressed this problem elegantly by adding supervised $F_0$ and energy as conditioning for its non-autoregressive decoder during training.
They are predicted in the same manner as duration and fed to the decoder during inference.
Yet another way to address this problem is to augment the model by incorporating an auxiliary representation to capture the latent factors in the speech data \cite{styletoken,hsu2018hierarchical,zhang-interspeech-2019,sun2020fullyhierarchical}.
An example of such latent factors in speech is prosody, which is isolated from linguistic content (text).
The auxiliary representation, which can be a form of style tokens \cite{styletoken} or a latent vector from a variational autoencoder (VAE), is extracted by an encoder taking groundtruth spectrogram as its input.

This paper presents a non-autoregressive neural TTS model augmented by a VAE.
The model, called \emph{Parallel Tacotron}\footnote{
Audio examples: \url{https://google.github.io/tacotron/publications/parallel_tacotron/}}, has the following properties which we found to be helpful to synthesize highly natural speech efficiently; 
(1) non-autoregressive architecture based on self-attention with lightweight convolutions \cite{wu2019pay},
(2) iterative mel-spectrogram loss \cite{dejavu},
(3) VAE-style residual encoder \cite{hsu2018hierarchical,sun2020fullyhierarchical}.

The rest of the paper is organized as follows.
Section~\ref{sec:architecture} explains the architecture of Parallel Tacotron.
Section~\ref{sec:experiments} describes experimental results.
Concluding remarks are given at the final section.

%% file: experiments.tex
\subsection{Training Setup}

 We used a proprietary speech dataset containing 405 hours of speech data; 347,872 utterances including 45 speakers in 3 English accents (32 US English speakers, 8 British English, and 5 Australian English speakers).
The Parallel Tacotron models were trained with Nesterov\footnote{Adam optimizer with the typical Transformer schedule performed worse.} momentum optimization and $\alpha=0.99$.  We used a linear warmup from 0.1--1.0 for the first 10K steps followed by the exponential decay from step 20K to 100K with a minimum value of 0.01. 
All models were trained for 120K steps with global gradient norm clipping of 0.2 and a batch size of 2,048 using Google Cloud TPUs.  Training took less than one day.

The fine-grained phoneme-level VAE model was trained differently. It used a KL-weight schedule where $\beta$ was increased linearly to 1.0 from step 6K to 50K.   Then the fine-grained phoneme-level VAE model was trained with exponential decay between steps 40K and 120K and we chose the checkpoint at step 215K.

As a baseline we used the Tacotron~2 model \cite{tacotron2} with some small modifications: specifically it used Gaussian mixture model (GMM) attention \cite{graves2013generating,transfer} and a reduction factor of 2 \cite{tacotron}, both of which have been shown to improve the robustness \cite{shen-arxiv-2020}. Tacotron~2 was trained for 500K steps with batch size 1,024 using Adam with exponential decay from step 50K to 400K.

Both baseline and proposed models were combined with the same pretrained WaveRNN neural vocoder \cite{kalchbrenner2018efficient} to reconstruct audio signals from predicted mel-spectrograms.

\subsection{Evaluation Setup}
Subjective evaluations were conducted using 1,000 sentences.
These sentences were different from the training data and was used in the previous papers.
They were synthesized using 10 US English speakers (5 male \& 5 female) in a round-robin style (100 sentences per speaker).
The naturalness of the synthesized speech was evaluated through subjective listening tests, including 5-scale Mean Opinion Score (MOS) tests and side-by-side preference tests. 
For MOS tests, a five-point Likert scale score (1: Bad, 2: Poor, 3: Fair, 4: Good, 5: Excellent) was adopted with rating increments of 0.5.
For side-by-side preference tests each rater listened to two samples then rated each with integral scores $[-3,3]$; where a positive score indicated that the first sample sounded better than the second one \cite{transfer, tacotron2}.

\subsection{Experimental Results}

\begin{table}[th]
\caption{Subjective evaluations of Parallel Tacotron with and without the iterative loss. Positive preference scores indicate that the corresponding Parallel Tacotron model was rated better than the reference Tacotron~2.}
\centering
\begin{tabular}{lrr}\\\toprule  
\textbf{Model} & \multicolumn{1}{c}{\textbf{MOS}} & \multicolumn{1}{c}{\textbf{Preference}}
\\\midrule
Tacotron~2& $4.45 \pm 0.04$ &  Reference \\
\midrule
\multicolumn{3}{l}{Parallel Tacotron (LConv, w/o VAE)}  \\
\quad single loss &  $4.32 \pm 0.04$& $\bf{-0.09 \pm 0.06}$\\
\quad iterative loss & $4.33 \pm 0.04$ & 	$\bf{-0.08 \pm 0.06}$  \\
\bottomrule
\end{tabular}
\label{evaluation_iter}
\vspace{1mm}
\caption{Subjective preference scores between Parallel Tacotron with and without the iterative loss.
Positive preference scores indicate that the corresponding model with the iterative loss was rated better than the one without the iterative loss.
}
\centering
\begin{tabular}{lr}\\\toprule  
\textbf{Model} &  \multicolumn{1}{c}{\textbf{Preference}}
\\\midrule
LConv w/o VAE  & $\bf{0.09 \pm 0.05}$\\
LConv w/ Global VAE &$\bf{0.07 \pm 0.05}$\\
Transformer w/ Global VAE & $\bf{0.08 \pm 0.05}$\\
\bottomrule
\end{tabular}
\label{preference_iter}
\vspace{1mm}
\caption{Subjective evaluations of Parallel Tacotron with different self-attention (with Global VAE and iterative loss).  Positive preference scores indicate that the corresponding Parallel Tacotron was rated better than Tacotron~2.}
\centering
\begin{tabular}{lrr}\\\toprule  
\textbf{Model} & \multicolumn{1}{c}{\textbf{MOS}} & \multicolumn{1}{c}{\textbf{Preference}}
\\\midrule
\multicolumn{3}{l}{Parallel Tacotron (w/ iterative loss \& Global VAE)}  \\
\quad Transformer  & $4.36 \pm 0.04$ & $-0.03 \pm 0.06$ \\
\quad LConv        & $4.40 \pm 0.04$ & $-0.01 \pm 0.05$ \\
\bottomrule
\end{tabular}
\label{evaluation_decoder}
\vspace{1mm}
\caption{Subjective preference score between Parallel Tacotron  using LConv and Transformer-based self-attention.
Positive preference scores indicate that LConv was rated better than Transformer.
}
\centering
\begin{tabular}{lrr}\\\toprule  
\textbf{Model} & \multicolumn{1}{c}{\textbf{Preference}}
\\\midrule
LConv \textit{vs} Transformer & $\bf{0.05 \pm 0.04}$\\
\bottomrule
\end{tabular}
\label{preference_decoder}
\end{table}

The first experiment evaluated the effect of the iterative loss. 
Tables~\ref{evaluation_iter} and \ref{preference_iter} show the experimental result.
Although there was no significant difference in MOS and preference against Tacotron~2, the direct comparison between models with and without the iterative loss indicate that it can give small improvement.

The second experiment evaluated the impact of VAEs.
Tables~\ref{evaluation_vae} and \ref{preference_vae} show the experimental results.
Parallel Tacotron without VAE was significantly worse than the baseline Tacotron~2 in both MOS and preference.
The introduction of global VAE made it comparable to the baseline Tacotron~2 in both evaluations.
Furthermore, the introduction of fine-grained phoneme-level VAE further boosted the naturalness.

\begin{table}[th]
\caption{Subjective evaluations of Parallel Tacotron with different VAEs.  Positive preference scores indicate that the corresponding Parallel Tacotron was rated better than Tacotron~2.}
\centering
\begin{tabular}{lrr}\\\toprule  
\textbf{Model} & \multicolumn{1}{c}{\textbf{MOS}} & \multicolumn{1}{c}{\textbf{Preference}}
\\\midrule
\multicolumn{3}{l}{Parallel Tacotron (w/ iterative loss \& LConv)}  \\
\quad No VAE     & $4.33 \pm 0.04$ & $\bf{-0.08 \pm 0.06}$  \\
\quad Global VAE & $4.40 \pm 0.04$ & $-0.01 \pm 0.05$ \\
\quad Fine VAE   & $4.42 \pm 0.04$ & $\bf{0.07 \pm 0.07}$ \\
\bottomrule
\end{tabular}
\label{evaluation_vae}
\vspace{1mm}
\caption{Subjective preference scores between Parallel Tacotron using the global and fine-grained VAEs. %
Positive preference scores indicate that left models were rated better than the right ones.
}
\centering
\begin{tabular}{lr}\\\toprule  
\textbf{Model} & \multicolumn{1}{c}{\textbf{Preference}}
\\\midrule
 Global VAE \textit{vs} No VAE & $\bf{0.13 \pm 0.05}$\\
 Fine VAE \textit{vs} Global VAE & $\bf{0.06 \pm 0.06}$\\
\bottomrule
\end{tabular}
\label{preference_vae}
\vspace{1mm}
\caption{Subjective preference scores between synthetic and natural speech.   The preference scores become positive when synthetic speech was rated better than natural one.}
\centering
\begin{tabular}{lr}\\\toprule  
\textbf{Model} & \multicolumn{1}{c}{\textbf{Preference}}
\\\midrule
Tacotron~2 &$\bf{-0.09 \pm 0.07}$\\ \midrule
\multicolumn{2}{l}{Parallel Tacotron w/ LConv \& iterative loss} \\
\quad Global VAE &  $-0.07 \pm 0.07$\\
\quad Fine VAE & $-0.01 \pm 0.06$\\
\bottomrule
\end{tabular}
\label{preference_human}
\vspace{1mm}
\caption{Inference speed to predict mel-spectrograms for $\sim$20-second long utterance on a TPU (aggregated over ten trials). }
\label{performance}
\centering
\begin{tabular}{lrr}\\\toprule  
\textbf{Model} & Mean \textbf{($1 / \textsc{RTF}$)} & Stddev 
\\\midrule
Tacotron~2& 79 & 1\\ \midrule
\multicolumn{2}{l}{Parallel Tacotron}  & \\
\quad Transformer w/ Global VAE & 804 & 5\\
\quad LConv w/ Global VAE & 1051 & 13\\
\quad LConv w/ Fine VAE & 1013 & 10\\
\bottomrule
\end{tabular}
\end{table}

The third experiment compared Transformer and LConv for self-attention.
Tables~\ref{evaluation_decoder} and \ref{preference_decoder} show the experimental results.
Parallel Tacotron with both Transformer and LConv-based self-attention matched the baseline Tacotron~2 both in MOS and preference.
When they were directly compared, Parallel Tacotron with LConv was more preferred.  

The last evaluation compared was done over 1,000 utterances from a held-out test set. 
This allowed us to make direct comparisons with natural speech. 
The MOS of natural speech was $4.54 \pm 0.04$ on this test set.
Experimental results are shown in Table~\ref{preference_human}.
It can be seen from the table that all neural TTS models are doing well compared to human speech, however there is still a room for further improvement.

\subsection{Inference Efficiency}
Table~\ref{performance} shows the inference time to predict mel-spectrograms for a 20 second-long utterance on a tensor processing unit (TPU). 
It can be seen from the table that Parallel Tacotron was about $13$ times faster than Tacotron 2. 
It also shows that the use of lightweight convolutions is faster than of Transformer.
It should be noted that little work has been done to trade-off model size and speed against the naturalness thus further speedup is highly possible.